\documentclass[acus]{JAC2003}
\usepackage{graphicx}
\usepackage{booktabs}
\usepackage{atlasphysics}

\begin{document}
\title{Determination of Integrated Luminosity via W and Z Boson Production with the ATLAS Detector}

\author{Matthias Schott, on behalf of the ATLAS Collaboration}

\maketitle

\begin{abstract}
The possibility to determine the recorded integrated luminosity via the measurements of the W and Z boson production cross-sections with the ATLAS detector is discussed. The current results based on 2010 data are briefly summarized. Special attention is drawn to theoretical uncertainties of the measurement. The latter give a large contribution to the systematic uncertainties of the measurements. An outlook on the expected precision of an analysis based on $\int L dt\approx 1\,\ifb$ is given and the implications on a possible luminosity determination are discussed.
\end{abstract}

\section{Theoretical Predictions}

The theoretical predicted cross-sections for the W and Z boson production at a p-p collider with $\sqrt{s}=7\TeV$, discussed in this note, are based on QCD NNLO calculations obtained with the FEWZ program \cite{FEWZ1, FEWZ2}. The central values and their corresponding theoretical uncertainties are summarized in Table \ref{tab:1}. The major sources of systematic uncertainties come from PDF and $\alpha_s$ measurements, based on a 90\% confidence limit, while electroweak radiative corrections have been assumed to be negligible. A detailed discussion can be found in \cite{ATLASWZ300}. 

\begin{table}[hbt]
   \centering
   \begin{tabular}{c|c}
	\hline
	        				& $\sigma_{tot} \cdot BR(W\rightarrow l\nu) [nb]$  \\ 
	\hline
       $W^+$			& $6.16 \pm 0.31$	\\ 
       $W^-$				& $4.30 \pm 0.21$ \\
       $W$				& $10.46 \pm 0.52$ \\
	\hline
        					& $\sigma_{tot} \cdot BR(Z/\gamma^{*}\rightarrow ll) [nb]$\\
					& $66<m_{ee} <116\GeV$  \\ 
	\hline
	$Z/\gamma^{*}$	& $0.964\pm0.048$\\
	\hline
   \end{tabular}
   \caption{Theoretical predictions of the W and Z boson production cross-sections based on QCD NNLO calculations by FEWZ.}
   \label{tab:1}
\end{table}

\section{Measurement of the W and Z Boson production cross-sections}

A detailed description of the preliminary measurements of the inclusive Drell-Yan $W\rightarrow l\nu$ and $Z/\gamma^*\rightarrow ll$ $(l=e,\mu)$ production cross-sections in proton-proton collisions at $\sqrt{s}=7\TeV$ can be found in \cite{ATLASWZ}. This study is based on the complete data collected in 2010, with an integrated luminosity of about 35$\,\ipb$. In the following, only a brief summary of the results is given.

The leptonic ($e,\mu$) decays of W and Z bosons provide clean final states to measure their production cross-sections in proton-proton collisions with the ATLAS detector. The detector and lepton identification strategy are described in \cite{ATLASDetector}.

The W and Z boson production cross-sections can be expressed as follows:

\begin{equation}
\label{Eqn:Cross}
\sigma_{W/Z} \times BR_{W/Z\rightarrow l\nu / ll} = \frac{N_S-N_B}{A_{W/Z} \cdot C_{W/Z} \cdot \int {\cal L} dt}
\end{equation}

where $N_S$ is the number of selected candidate events in data, $N_B$ is the number of estimated background events and $A_{W/Z}$ corresponds to the geometrical and kinematic acceptance for the W/Z boson decays under consideration at purely theoretical level. The factors $C_{W/Z}$ allow to correct the number of observed events for detector effects in the acceptance region. This includes all reconstruction-, trigger- and cut-efficiencies and $\int {\cal L} dt$ denotes the integrated luminosity for the channel of interest. 

The lepton trigger, reconstruction and identification efficiencies have been estimated in data using the so-called 'tag-and-probe' technique applied on samples of Z bosons decaying into two leptons \cite{TagAndProbe}. In addition, the lepton momentum scale and resolution as well as effects on the $\met$ scale and resolution have been studied with data-driven approaches \cite{ATLASWZ}. The experimental correction factors $C_W$ and $C_Z$ are determined via Monte Carlo simulations of the ATLAS detector and have been corrected for discrepancies between data and Monte Carlo. 

The fiducial regions for the W and Z boson production cross-sections are defined via the following cuts applied at the generator level for the electron and muon decay channels, respectively:

\begin{itemize}
\item $W\rightarrow e \nu$: $\et^e >20\GeV$, $|\eta|<2.47$, excluding $1.37<\eta<1.52$, $\et^{\nu} >25\GeV$, $m_T > 40\GeV$
\item $W\rightarrow \mu \nu$: $\pt^\mu >20\GeV$, $|\eta|<2.4$, $\et^{\nu} >25\GeV$, $m_T > 40\GeV$
\item \Zee: $\et^e >20\GeV$, $|\eta|<2.47$, excluding $1.37<\eta<1.52$, $66\GeV< m_{ee} < 106\GeV$
\item \Zmm: $\pt^\mu >20\GeV$, $|\eta|<2.4$, $66\GeV < m_{\mu\mu} < 106\GeV$
\end{itemize}

The purely theoretical factor $A_{W/Z}$ is used to extrapolate from the experimentally accessible fiducial region to the full phase-space and has been determined on Monte Carlo generator level. The systematic uncertainties on $A_{W/Z}$ vary between 3\% and 4\% and are dominated by PDF uncertainties. The theoretical uncertainties on the detector and efficiency correction factors $C_{W/Z}$ can be assumed to be negligible to a good extend.

\begin{figure}[htb]
   \centering
   \includegraphics*[width=80mm]{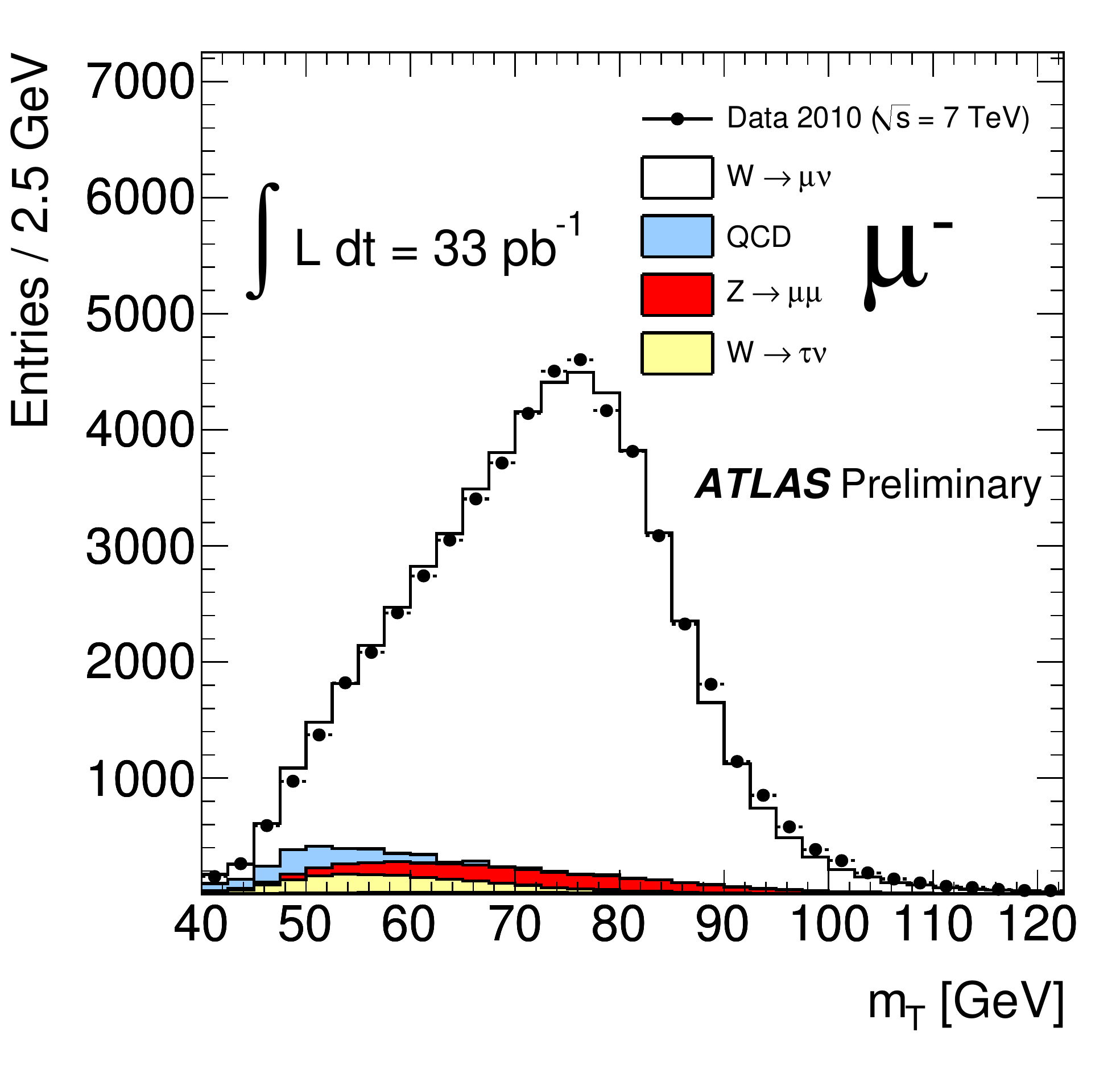}
   \caption{Transeverse mass $m_T$ distribution for $W^{-}\rightarrow \mu^{-}\nu$ candidates selected in the full 2010 data-set.}
   \label{Fig:1}
\end{figure}

The recorded electron and muon data samples, used for this analysis, were selected by single lepton triggers, using a combination of hardware and software based triggers. The selection of W boson candidate events requires one identified lepton with a large transverse momentum ($\pt$) above 20$\GeV$ and within the geometrical acceptance of the detector as already introduced above. Muons are reconstructed in the ATLAS inner detector and muon spectrometer which give rise to two independent $\pt$ measurements. The selected muon candidates are required to be isolated and have a good matching between the muon spectrometer and the corresponding inner detector measurement. The electron candidates have to pass several cuts on their shower shapes in the electromagentic calorimeter, their inner detector track properties and the track-cluster matching between the ATLAS inner detector and the calorimeter system. In addition, it is required that the missing transverse energy ($\met$) of the event is larger than 25$\GeV$ and that the reconstructed transverse mass ($m_T$) defined by the lepton-$\met$ system, is larger than 40$\GeV$. Figure \ref{Fig:1} shows the selected $m_T$ distribution for negative charged W boson candidates in the muon decay channel. The QCD background is estimated by data-driven methods on both the electron and muon channels, while the electroweak contribution is obtained from Monte Carlo simulations.

\begin{figure}[htb]
   \centering
   \includegraphics*[width=80mm]{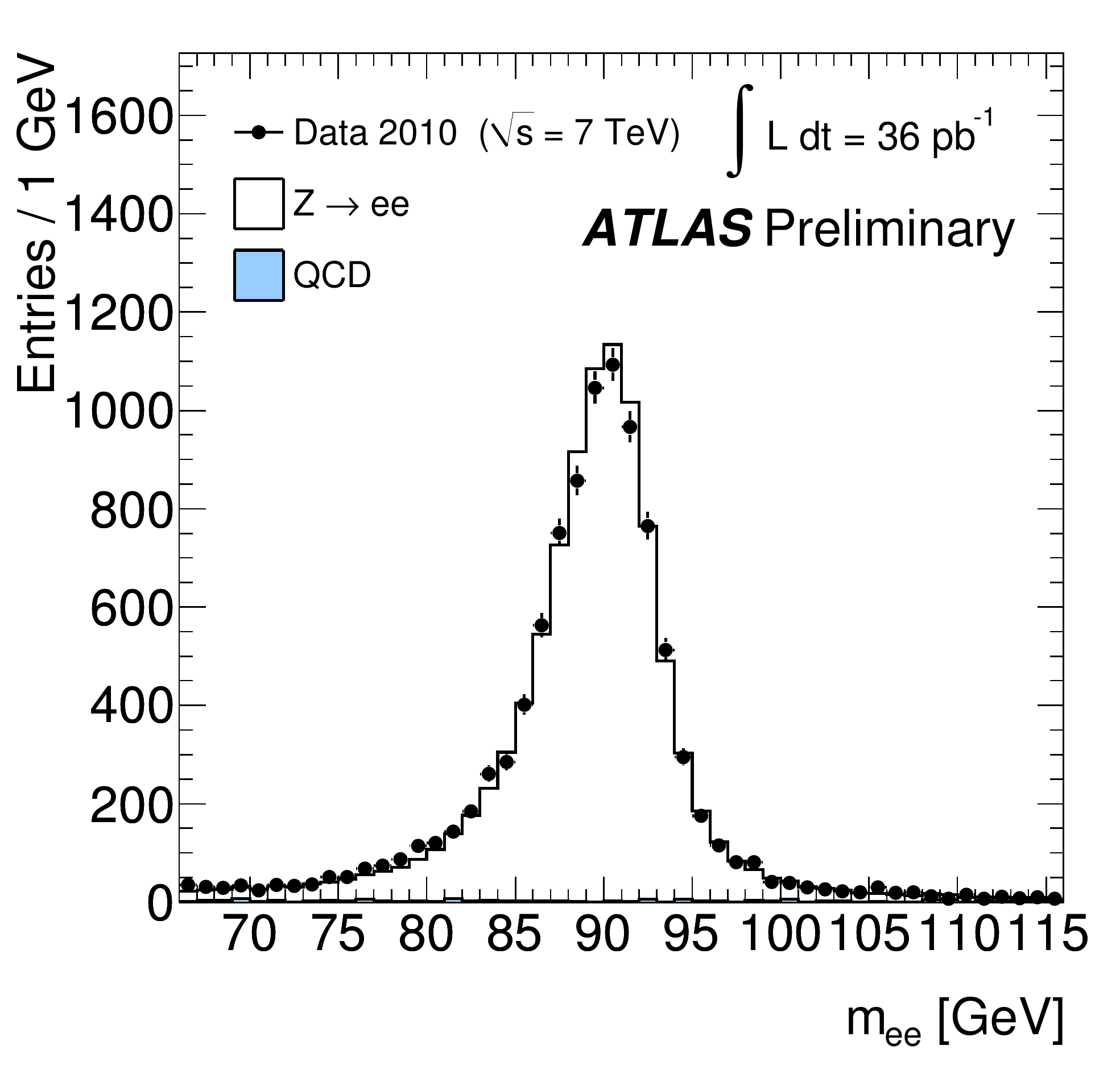}
   \caption{Invariant mass distribution for \Zee candidates selected in the full 2010 data-set.}
   \label{Fig:2}
\end{figure}

The selection of \Zll candidate events requires two oppositely charged leptons with a large transverse momentum $\pt>20\GeV$. The reconstructed muons must satisfy the same selection criteria and isolation cuts as in the W boson selection. The electron candidates have to pass slightly looser quality criteria. The invariant di-lepton mass $m_{ll}$ is required to be within $66\GeV<m_{ll}<116\GeV$ for both decay channels. Figure \ref{Fig:2} shows the resulting invariant mass distribution of the selected electron pairs in comparison with the Monte Carlo expectation.

In total, 121,310 $W^\pm\rightarrow e^\pm\nu$, 9,721 $\Zee$, 139,266 $W^\pm\rightarrow \mu^\pm\nu$ and 11,669 \Zmm candidate events have been selected in the complete 2010 data-set, corresponding to an integrated luminosity of about 35$\ipb$. The measured cross-sections for the combination of the electron and muon decay channels are summarized in Table \ref{Tab:2} together with the statistical and systematic uncertainties. The luminosity uncertainty as well as the uncertainty of the acceptance corrections are shown separately. The experimental uncertainties on the Z boson cross-section is $1.2\%$, excluding the luminosity uncertainty. This is very close to the ATLAS predictions based on studies at $\sqrt{s}=14\TeV$ with an assumed integrated luminosity of 50$\,\ipb$ \cite{ATLASCSC}.

\begin{table}[hbt]
   \centering
\begin{tabular}{c|c|c|c|c|c}
\hline
 &\multicolumn{5}{c}{$\sigma_{tot} \cdot BR(W\rightarrow l\nu) $}\\
\cline{2-6}
 &Value 			& \stat	& \syst	& \lumi	& \acc	\\
 & [nb] 			& [nb]	& [nb]	& [nb]	& [nb]	\\
\hline\hline
$W^+$ 			&	6.257	& 0.017 	& 0.152	& 0.213	& 0.188	\\
$W^-$ 			&	4.149	& 0.014 	& 0.102	& 0.141	& 0.124	\\
$W^\pm$ 			&	10.391	& 0.022 	& 0.238	& 0.353	& 0.312	\\
\hline
$Z/\gamma^{*}$ 	&	0.945	& 0.006 	& 0.011	& 0.032	& 0.038	\\
\hline
\end{tabular}
   \caption{Total averaged cross-sections times leptonic branching ratios for $W^+$, $W^-$, $W$ and $Z/\gamma^*$ production in the combined electron and muon final states. The uncertainties denote the statistical \stat, experimental systematic \syst errors, the luminosity induced \lumi errors as well as the uncertainty of the acceptance factor $A_{W/Z}$ \acc in Equation \ref{Eqn:Cross}.
}
   \label{Tab:2}
\end{table}

\begin{figure}[htb]
   \centering
   \includegraphics*[width=80mm]{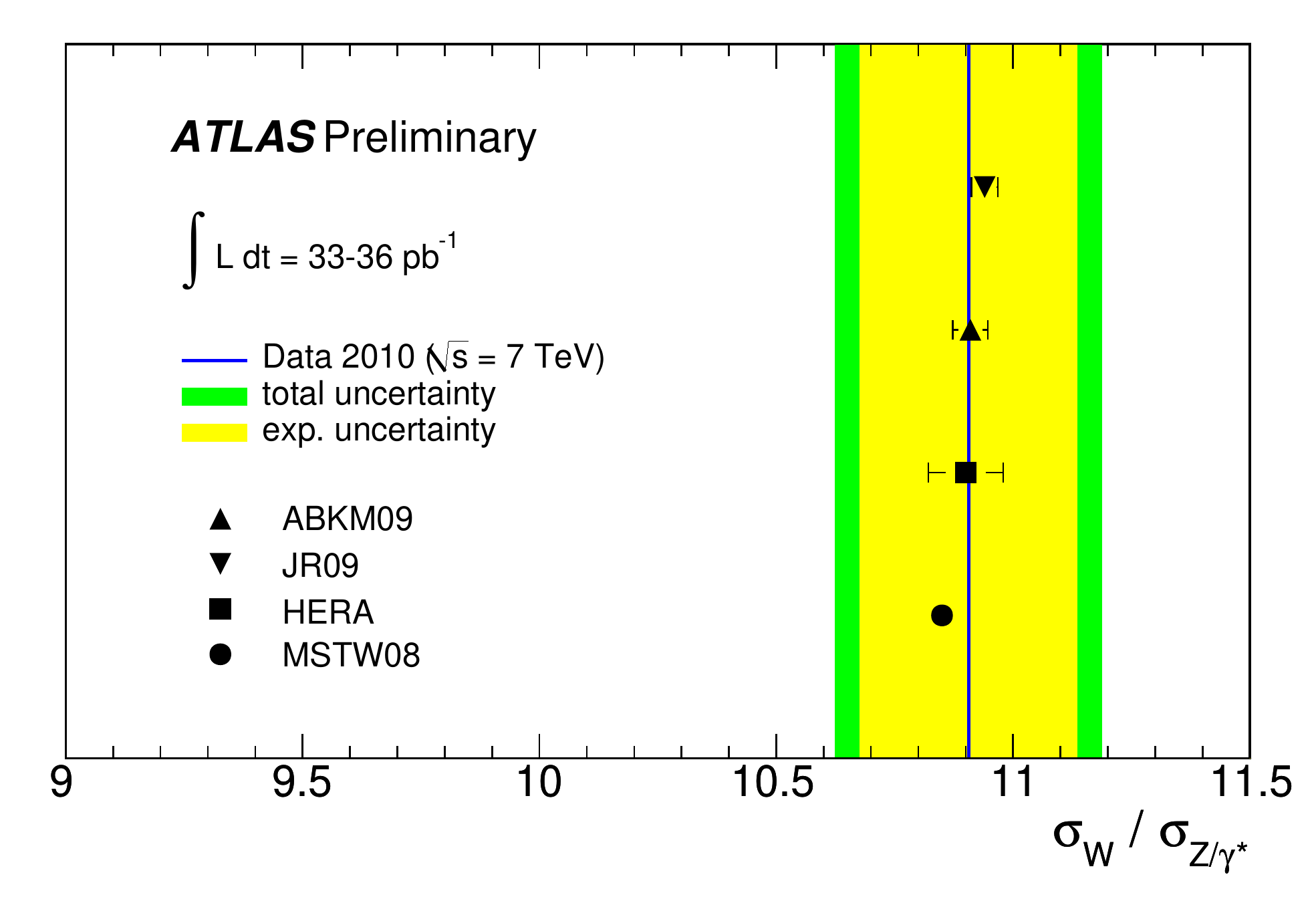}
   \caption{Predicted cross-section ratio $\sigma_W/\sigma_{Z/\gamma^*}$ and the corresponding measurement: 10.906 $\pm$ 0.079\stat $\pm$ 0.215\syst $\pm$ 0.164$\acc$ \cite{ATLASWZ}. The experimental uncertainty of the measurement includes statistical and experimental systematic errors. The uncertainties of the predicted cross-section ratios are very small.}
   \label{FigRatio}
\end{figure}

The results are in very good agreement with the theoretical calculations at NNLO in QCD, as summarized in Table \ref{tab:1}. A powerful test of NNLO QCD calculations is provided by the ratio of W and Z boson production cross-sections, since several theoretical and experimental systematic uncertainties cancel to a good extend in the ratio measurement. The comparison between the measured and predicted ratio is presented in Figure \ref{FigRatio} which clearly highlights the large agreement between experiment and theory.

\section{Luminosity Determination based on Vector Boson Production}

Equation \ref{Eqn:Cross} can be rearranged to predict the integrated luminosity of the studied data-sample, by assuming that the theoretically predicted cross-sections agrees with the experimental measurements. In the following it will be discussed which precision on the integrated luminosity can be expected by studying the W and Z boson production at LHC. After the data-taking of roughly 300k Z boson events per lepton decay channel (corresponding to $\int {\cal L} dt=1\,\ifb$) the statistical component of the measurement is no longer critical in any respect. In addition, data-driven detector performance studies can improve the understanding of the detector performance, leading to a relative uncertainty below $1\%$ on the detector correction factors $C_{W/Z}$ in Equation \ref{Eqn:Cross}. Hence the dominant systematic in the final luminosity estimate will be the theoretical uncertainty on $A_{W/Z}$, which are currently of the order of 3-4\%. This large uncertainty comes mainly from the extrapolation of $\eta$ beyond the fiducial region, while the extrapolation to lower values of $\pt$ has only a minor effect. 

One possibility to lower the uncertainty on $A_{W/Z}$ is to increase the fiducial volume of the measurement by including the ATLAS electron identification up to  $|\eta|<4.5$ in the full cross-section measurement. This would reduce the required extrapolation and thus the systematic uncertainty on $A_{W/Z}$. This has been investigated in \cite{ATLASWZ}. In addition, a differential measurement of the production cross-sections of W and Z bosons as a function of $\pt$ and $\eta$ can be used to constrain PDF sets and reduce the corresponding systematic uncertainties significantly. One step in this direction was the measurement of the $W^{+}/W^{-}$ charge asymmetry, shown in Figure \ref{FigWAsym} and discussed in detail in \cite{ATLASWPMPaper}. 

\begin{figure}[htb]
   \centering
   \includegraphics*[angle=90,height=80mm]{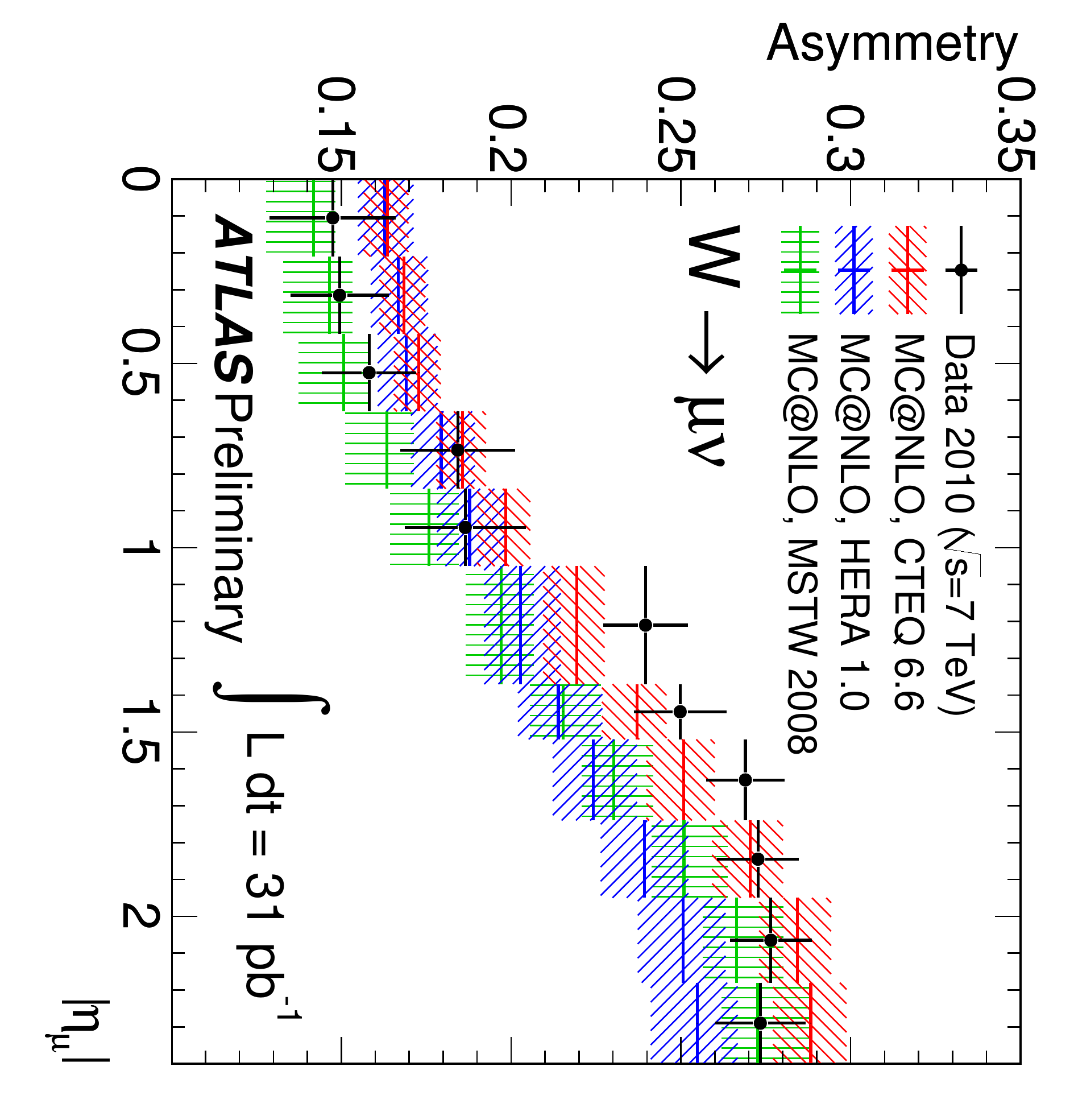}
   \caption{The muon charge asymmetry from W boson decays in bins of absolute pseudorapidity including statistical and systematic contributions. The kinematic requirements applied are $p^{\mu}_T >20\GeV$, $p^{\nu}_T >25\GeV$ and $m_T>40\GeV$.}
   \label{FigWAsym}
\end{figure}

Another approach would be to avoid uncertainties on $A_{W/Z}$ at a theoretical level by comparing measured fiducial cross-sections with theoretical predictions in the fiducial region. New versions of NNLO cross-section programs allow the calculation of fiducial cross-sections, i.e. cross-sections within the detector acceptance \cite{FEWZ2}. These programs have currently no parton show model nor resummation effects included and hence lead to an unrealistic $\pt$ distribution of the decay leptons. Future studies will show how this situation can be improved and uncertainties due to these missing effects can be estimated. 

\section{Conclusion}

The lower bound on the theoretical uncertainty on the W and Z boson production cross-section predictions is given by renormalisation and factorisation scale uncertainties (0.6\%-0.8\%) which cannot be lowered without significant improvements on the higher order QCD calculations. The current theoretical uncertainties for cross-section predictions in the fiducial region of the detector are estimated at the $\approx3\%$ level, dominated by PDF uncertainties \cite{Alekin}. Those are expected to be significantly lowered in the coming years by future precision measurements at the LHC while a quantitative statement on the expected uncertainty reduction cannot be made. The expected experimental uncertainties are expected to be significantly below $1\%$. The expected precision on the integrated luminosity determination via W and Z boson production at the ATLAS detector is therefore expected to be $1-3\%$.

However, it should be noted that such an approach essentially relates all cross-section measurements of further physics processes (e.g. the top-quark pair production), to the production cross-section of the electroweak bosons. Hence, production cross-section measurements are essentially replaced by the measurement of cross-section ratios.


\begin{thebibliography}{9}   

\bibitem{FEWZ1} C. Anastasiou, L. J. Dixon, K. Melnikov, and F. Petriello, High precision QCD at hadron colliders: Electroweak gauge boson rapidity distributions at NNLO, Phys. Rev. D69 (2004) 094008, arXiv:hep-ph/0312266

\bibitem{FEWZ2} R. Gavin, Y. Li, F. Petriello, and S. Quackenbush, FEWZ 2.0: A code for hadronic Z production at next-to- next-to-leading order, arXiv:1011.3540 [hep-ph]

\bibitem{ATLASWZ300} ATLAS Collaboration, Measurement of the W and Z production cross-sections in proton-proton collisions at $\sqrt{s}=7\TeV$ with the ATLAS detector, JHEP 12 (2010) 060, arXiv:1010.2130 [hep-ex]

\bibitem{ATLASWZ} ATLAS Collaboration, A measurement of the total W and Z/$\gamma^*$ cross-sections in the e and $\mu$ decay channels and of their ratios in pp collisions at $\sqrt{s}=7\,TeV$ with the ATLAS detector, ATLAS-CONF-2011-041

\bibitem{ATLASDetector} ATLAS Collaboration, The ATLAS Experiment at the CERN Large Hadron Collider, JINST 3 (2008) S08003

\bibitem{TagAndProbe} ATLAS Collaboration, Muon Performance Note, Phys. Rev. Lett. 25 (1997) 56
ATLAS Collaboration, Determination of the muon reconstruction efficiency in ATLAS at the Z resonance in proton-proton collisions at $\sqrt{s}=7\TeV$, ATLAS-CONF-2011-008

\bibitem{ATLASCSC}
The ATLAS Collaboration, Expected Performance of the ATLAS Experiment - Detector, Trigger and Physics, arXiv:0901.0512 [hep-ex]
  
\bibitem{ATLASWPMPaper} ATLAS Collaboration, Measurement of the Muon Charge Asymmetry from W Bosons Produced in pp Collisions at $\sqrt{s}=7\TeV$ with the ATLAS detector, arXiv:1103.2929; CERN-PH-EP-2011-036

\bibitem{Alekin} S. Alekhin, J. BlŸmlein, P. Jimenez-Delgado, S. Moch, E. Reya, NNLO Benchmarks for Gauge and Higgs Boson Production at TeV Hadron Colliders, arXiv:1011.6259; 2010


\end{thebibliography}
\end{document}